\documentclass[preprint,5p]{elsarticle}

\usepackage{graphicx}

\newcommand{\insertplot}[5]{\begin{figure}
 \hfill\hbox to 0.05in{\vbox to #5in{\vfill
 \inputplot{#1}{#4}{#5}}\hfill}
 \hfill\vspace{-.1in}
 \caption{#2}\label{#3}
 \end{figure}}
 \newcommand{\inputplot}[3]{
 \special{ps: plotfile #1}
}
\newcounter{fig}   

\newcommand{\vphi}{\varphi}

\usepackage{graphicx}

\begin{document}

\title{
Angular momentum -- area -- proportionality\\ of extremal charged black holes
in odd dimensions}

\vspace{1.5truecm}
\author{
{\bf Jose Luis Bl\'azquez-Salcedo$^1$},
{\bf Jutta Kunz$^2$},
{\bf Francisco Navarro-L\'erida$^3$}\\
%
%
$^1$
Dept.~de F\'{\i}sica Te\'orica II, Ciencias F\'{\i}sicas\\
Universidad Complutense de Madrid, E-28040 Madrid, Spain\\
$^2$ Institut f\"ur  Physik, Universit\"at Oldenburg\\ Postfach 2503,
D-26111 Oldenburg, Germany\\
$^3$
Dept.~de F\'{\i}sica At\'omica, Molecular y Nuclear, Ciencias F\'{\i}sicas\\
Universidad Complutense de Madrid, E-28040 Madrid, Spain
}


\date{\today}


\begin{abstract}
Extremal rotating black holes in Einstein-Maxwell theory
feature two branches. On the branch emerging from
the Myers-Perry solutions their angular momentum is proportional
to their horizon area, while on the branch emerging
from the Tangherlini solutions their angular momentum is proportional
to their horizon angular momentum. The transition between these branches
occurs at a critical value of the charge, which depends on the value of the angular momentum.
However, when a dilaton is included, the angular momentum
is always proportional to the horizon area.
\end{abstract}

\maketitle



\section{Introduction}

Although in $D=4$ dimensions the Kerr-Newman solution represents the
unique family of stationary asymptotically flat black holes 
of Einstein-Maxwell (EM) theory, the corresponding $D>4$ charged
rotating black holes have not been obtained in closed form yet. Only certain
subsets are known: the generalization of the static black hole to higher
dimensions pioneered by Tangherlini 
\cite{Tangherlini:1963bw}, and the rotating vacuum black holes, obtained by
Myers and Perry (MP) \cite{Myers:1986un}. Other subsets
could be constructed perturbatively
\cite{Aliev:2004ec,Aliev:2006yk,NavarroLerida:2007ez,Allahverdizadeh:2010xx,Allahverdizadeh:2010fn}
and numerically \cite{Kunz:2005nm,Kunz:2006eh}.\footnote{In 
this paper we will consider
only asymptotically flat black holes 
with spherical horizon topology.}

Nevertheless, if additional fields and/or interactions are allowed into the
theory, exact higher dimensional charged rotating black holes can be obtained
by solution generating techniques. For example, in the simplest Kaluza-Klein (KK) case,
a boost is done to the $D+1$ embedding of the uncharged $D$-dimensional MP
black holes along the extra dimension. The result is a charged $D$-dimensional
black hole in Einstein-Maxwell-dilaton (EMd) theory. The dilaton coupling
constant $h$ for this solution has a 
particular value, which we denote $h_{\rm KK}$, that depends on the dimension
$D$ \cite{Kunz:2006jd}. 
To generate rotating EMd black hole solutions with other values of the coupling
constant $h$, 
currently perturbative or numerical techniques must be used.

$D$-dimensional stationary black holes possess, in general, $N$
independent angular momentum $J_i$  
associated with $N$ orthogonal planes of rotation \cite{Myers:1986un}, where
$N$ is the integer part of $(D-1)/2$, 
corresponding to the rank of the rotation group
$SO(D-1)$. As a result, we can distinguish between odd-$D$ and even-$D$ black
holes, where the latter have an unpaired spatial coordinate
\cite{Myers:1986un}. In the particular case in which all $N$ angular momenta
are equal in magnitude, the EMd equations simplify considerably, 
yielding, for odd dimensions, cohomogeneity-1 equations from which the angular
dependence can be extracted analytically. Hence, the equations
reduce to a more tractable system of ordinary differential equations.

When the $N$ angular momenta are of equal magnitude, $J= |J_i|$, it is
interesting to note that, for extremal MP black
holes, the angular momentum $J$ and the horizon area $A_{\rm H}$ are proportional:
$J = \sqrt{2(D-3)} A_{\rm H}$. 
This is a special case of a more general type of relations for
MP black holes in terms of the non-degenerate inner and outer horizon areas 
of non-extremal black holes
\cite{Cvetic:2010mn}, and was pointed out in 4 dimensions before
\cite{Ansorg:2007fh,Hennig:2008yw,Ansorg:2008bv,Hennig:2008zy,Ansorg:2009yi,Hennig:2009aa,Ansorg:2010ru}.    
In the case of charged black holes, the relation for the product of the
horizon areas can be typically written as a sum between the squares of the
angular momentum and some power of the charge
\cite{Cvetic:2010mn,Ansorg:2007fh,Hennig:2008yw,Ansorg:2008bv,Hennig:2008zy,Ansorg:2009yi,Hennig:2009aa,Ansorg:2010ru,Castro:2012av}.   

In this paper we study this kind of relations between area, angular
momentum and charge for extremal EM and EMd black holes with equal angular 
momentum. We construct the global solutions numerically and 
local solutions in the
near horizon formalism. The EM case is special since two different branches of
charged extremal solutions exist. One branch emerges from the uncharged MP
black holes, and the other branch emerges from the static Tangherlini black
holes. The area relations are different on each branch: the first branch
retains the proportionality 
between the angular momentum and the area of the MP solutions.
Thus the area of these charged black holes is independent of the charge.
In constrast, the second branch exhibits
a proportionality between the angular momentum
and the horizon angular momentum, while the charge enters into the area relation
yielding $A_{\rm H}^2 = C_1 J^2 Q^{-3/2} + C_2 Q^{3/2}$, where $C_1$ and $C_2$
are some constants and $Q$ is the electric charge.
However, as soon as the dilaton is coupled, the branch structure
changes, and only a single branch -
similar to the first branch of the EM case - is
found. Again along this branch the proportionality
between the angular momentum and the area persists for
all extremal solutions. We will proceed by first presenting the $D=5$ results
and then discussing their generalization to odd $D>5$ dimensions.

\section{5D EMd near horizon solutions}

In $5$ dimensions, the EMd action can be written as 
\begin{eqnarray} \label{EMDac}
&& I= \int d^5x \sqrt{-g} {\cal L} = \\ &&\int d^5x\sqrt{-g} \biggl[ R
 -\frac{1}{2}\partial_\mu\phi \, \partial^\mu\phi
 -\frac{1}{4}e^{-2h\phi}F_{\mu\nu}F^{\mu\nu} \biggr ] \ ,  \nonumber
\end{eqnarray}
where $R$ is the curvature scalar, $\phi$ the scalar dilaton field, $h$ the
dilaton coupling constant and $ F_{\mu \nu} = \partial_\mu A_\nu -\partial_\nu
A_\mu $ the field strength tensor, where $A_\mu $ denotes the gauge vector
potential. The units have been chosen so that $16 \pi G =1$, $G$ being
Newton's constant. If we set $h=0$, the pure EM action is recovered, while
$h_{\rm KK} =\sqrt{ \frac{2}{3} }$ is the KK value.

For cohomogeneity-1 solutions the
isometry group is enhanced from $\mathcal{R} \times U(1)^{2}$
to $\mathcal{R} \times U(2)$, where $\mathcal{R}$ represents time translations.
This symmetry enhancement allows to factorize the angular dependence
and thus leads to ordinary differential equations.

Following the near horizon formalism
\cite{Astefanesei:2006dd,Goldstein:2007km}, we now obtain exact near
horizon solutions for these extremal EM and EMd black holes. In terms of the
left-invariant 1-forms 
$\sigma_1=\cos \psi d\bar \theta+\sin\psi \sin  \bar\theta d \phi$,
$\sigma_2=-\sin \psi d\bar \theta+\cos\psi \sin  \bar\theta d \phi$, and
$\sigma_3=d\psi  + \cos  \bar\theta d \phi$, the near horizon metric can be
written as \begin{eqnarray}
&& ds^2=v_1(\frac{dr^2}{r^2}-r^2 dt^2) \nonumber \\ &&  +\frac{v_2}{4}(\sigma_1^2+\sigma_2^2)
+ \frac{v_2v_3}{4}(\sigma_3+2 k r dt)^2,
\label{at1}
\end{eqnarray}
where we have defined  $2\theta=\bar \theta$, $\phi_2-\phi_1=\phi$,
$\phi_1+\phi_2=\psi$, 
$\theta  \in [0,\pi/2]$, $(\varphi_1,\varphi_2) \in [0,2\pi]$.
The horizon is located at $r=0$, which can always be achieved via
a transformation $r\to r-r_H$. Note, that the metric is written in a
co-rotating frame.

The metric corresponds to a rotating squashed $AdS_2\times S^3$ spacetime,
describing the neighborhood of the event horizon of an extremal black hole. 
The corresponding Ansatz for the gauge potential in the
co-rotating frame reads
\begin{eqnarray}
\label{gv1}
&& A_\mu d x^\mu =
q_1 r dt + q_2 \sin^2 \theta \left( d\varphi_1 - k r  dt \right)
\nonumber \\     &&    + q_2 \cos^2 \theta \left( d\varphi_2 - k r  dt \right) \ .
\end{eqnarray}
The dilaton field is simply given by $\Phi=u$.
The parameters $k$, $v_i$, $q_i$ and $u$ are constants, and satisfy a set of
algebraic relations, which can be obtained, according to 
\cite{Astefanesei:2006dd,Goldstein:2007km}, in the following way.

Evaluating the Lagrangian density $\sqrt{-g} {\cal L}$
for the near horizon geometry (\ref{at1})
and integrating over the angular coordinates yields the function $f$,
\begin{equation}
\label{atd2}
f(k,v_1,v_2,v_3,q_1,q_2,u)
=\int d \theta d \varphi_1 d\varphi_2 \sqrt{-g} {\cal L} \ , 
\end{equation}
from which the field equations follow. 
In particular, the derivatives of $f$ with respect to the parameters
vanish except for
\begin{eqnarray}
\label{at3}
 \frac{\partial f}{\partial k}   = 2J \ , \ \ \
 \frac{\partial f}{\partial q_1} = Q \ ,
\end{eqnarray}
where $J$ is the total angualar momentum and $Q$ is the charge. From these
equations  a set of
algebraic relations for the near horizon expressions (\ref{at1}), (\ref{gv1}) is obtained.

The entropy function is obtained by taking the
Legendre transform of the above integral with respect to the
the parameter $k$, associated with both
angular momenta, $J_1=J_2=J$,
and with respect to the parameter $q_1$, associated with the charge $Q$,
\begin{eqnarray}
\label{at4}
&& {\cal E}(J,k,Q,q_1,q_2,v_1,v_2,v_3,u)=\nonumber \\ && 2 \pi \left(2 J k + Q q_1 -
f(k,v_1,v_2,v_3,q_1,q_2,u)\right).
\end{eqnarray}
Then the entropy associated with the black holes can be calculated by
evaluating this function at the extremum, $S={\cal E}_{extremal}$.

For the discussion of the solutions we need to consider
the EM and the EMd case separately.
In the EM case, i.e. for $h=0$,
the system of equations yields two distinct solutions,
depending on two parameters. These two solutions of the near horizon geometry
have been found independently by Kunduri and Lucietti in
\cite{Kunduri:2013gce}.\footnote{We thank Hari Kunduri for pointing this
out to us.}
Here we now calculate the charges and entropies
associated
with these two branches.
The solution containing the MP limit, and thus the first branch,
has $v_2=4 v_1$, $v_3=2 -\frac{q_2^2}{v_1}$, $q_1=0$, $k=\frac{1}{2}$ and
\begin{eqnarray}
&& J = 32 \pi^2 v_1 \sqrt{2 v_1 - q_2^2}  \ , \ \ \
Q = - 32 \pi^2 q_2 \sqrt{2 v_1 - q_2^2} , \nonumber \\
&& S = 2 \pi J \ , \ \ \
J_H = 16\pi^2(2v_1-q_2^2)^{3/2} \ ,
\end{eqnarray}
while the solution containing the Tangherlini limit, and thus the second branch,
has $v_2=4 v_1$, $v_3 = \frac{1}{4 k^2+1}$,
$q_1 = -\frac{(2k+1)(2k-1)\sqrt{3}}{2} \sqrt{ \frac{|v_1|}{4k^2+1}}$,
$q_2 = -2\sqrt{3}k \sqrt{ \frac{|v_1|}{4k^2+1}}$ and
\begin{eqnarray}
&& J = 128 \pi^2 k \left( \frac{|v_1|}{4k^2+1} \right)^{3/2} , 
Q = 32 \sqrt{3} \pi^2   \frac{v_1}{4k^2+1} , \nonumber \\
&& S= 64 \pi^3 {\frac{|v_1|^{3/2}}{\sqrt{4k^2+1}}} , 
J_H = 
 J/4,
\end{eqnarray}
where $J_{\rm H}$ is the horizon angular momentum.
The two solutions match at $k=1/2$, where $q_1=0$.
At this critical point the angular momentum can be written as
\begin{equation}
J = \frac{1}{2\sqrt{2}\pi}\frac{1}{3^{3/4}}Q^{3/2}.
\end{equation}
Thus we have the surprising result that
along the first branch, 
the proportionality of the angular momentum and the area
known for the MP black holes,
continues to hold in the presence of charge until the critical point is reached.
In contrast, on the second branch we have proportionality of the 
angular momentum and the horizon angular momentum.

In the case of the EMd black holes,
only one solution is found. It can be obtained by replacing ${q_2} \to
q_2e^{-hu},\bar{Q} = Qe^{hu} $ in the first branch solution of the EM case.
Hence, as long as $h \ne 0$, the angular momentum and the area are always proportional,
independent of $h$ and $Q$.
In particular, this includes the KK case, where the full solution is known analytically.

\section{5D EMd black hole solutions}

We now need to consider the full solutions, which we obtain by
numerical integration.
For the metric we employ the parametrization
\begin{eqnarray}
\label{metric}
&&ds^2 = -f dt^2 + \frac{m}{f}(dr^2 + r^2 d\theta^2) 
\\
\nonumber
&&
+ 
\frac{n}{f}r^2 \sin^2\theta \left( d \varphi -\frac{w}{r}dt \right)^2 
\\
\nonumber
&&+\frac{n}{f}r^2 \cos^2\theta \left( d \psi -\frac{w}{r}dt \right)^2 
\\
\nonumber
&&+\frac{m-n}{f}r^2 \sin^2\theta \cos^2\theta(d \varphi -d \psi)^2 \ ,
\end{eqnarray}
for the gauge potential we use
\begin{equation}
A_\mu dx^\mu = a_0 dt + a_k (\sin^2 \theta d\varphi+\cos^2 \theta d\psi) \ ,
\end{equation}
while the dilaton field is described by the function $\Phi(r)$.

The resulting set of coupled ODEs then consists of
first order differential equations for $a_0$ and $n$, and second order
differential equations for $f$, $m$, $n$, $\omega$, $a_k$ and
$\Phi$. The equation for $a_0$ allows to eliminate 
this function from the system.

To obtain asymptotically flat solutions, the metric functions should satisfy
the following set of boundary conditions 
at infinity,
$f|_{r=\infty}=m|_{r=\infty}=n|_{r=\infty}=1$, 
$\omega|_{r=\infty}=0$.
For the gauge potential we choose a gauge such that
$a_0|_{r=\infty}=a_\vphi|_{r=\infty}=0$.
For the dilaton field we choose
$\phi|_{r=\infty}=0$,
since we can always make a transformation $\phi \rightarrow \phi - \phi|_{r=\infty}$.

In isotropic coordinates the horizon is located at $r_{\rm H}=0$.
An expansion at the horizon yields
$f(r) = f_4 r^4 + f_{\alpha} r^{\alpha} + o(r^6)$,
$m(r) = m_2 r^2 + m_{\beta} r^{\beta} + o(r^4)$,
$n(r) = n_2 r^2 + n_{\gamma} r^{\gamma} + o(r^4)$,
$\omega(r) = \omega_1 r + \omega_2 r^2 + o(r^3)$,
$a_0(r) = a_{0,0} + a_{0,\lambda} r^{\lambda} + o(r^2)$,
$a_k(r) = a_{k,0} + a_{k,\mu} r^{\mu} + o(r^2)$,
$\Phi(r) = \Phi_0 + \Phi_{\nu} r^{\nu} + o(r^2)$.
Interestingly, the coefficients $\alpha$, $\beta$, $\gamma$,
$\lambda$, $\mu$ and $\nu$ are non-integer.
Only $\omega$ has an integer expansion.

To construct the solutions numerically,
we employ a compactified radial coordinate, $x= r/(r+1)$.
We then reparametrize the metric in terms of the functions
$f = \hat f x^2$,
$m = \hat m$,
$n(r) = \hat n$,
$\omega(r) = \hat \omega (1-x)^2$,
$a_k = \hat a_{k}/x^2$,
and
$\Phi = \hat \Phi/x^2$
to properly deal with the non-integer coefficients in the horizon expansion,
eliminating possible divergences in the integration of the functions.

We employ a collocation method for boundary-value ordinary
differential equations, equipped with an adaptive mesh selection procedure
\cite{COLSYS}.
Typical mesh sizes include $10^3-10^4$ points.
The solutions have a relative accuracy of $10^{-10}$.
The estimates of the relative errors of the global charges
and other physical quantities are of order $10^{-6}$.

\begin{figure}[t!]
\begin{center}
\includegraphics[height=.35\textheight, angle =270]{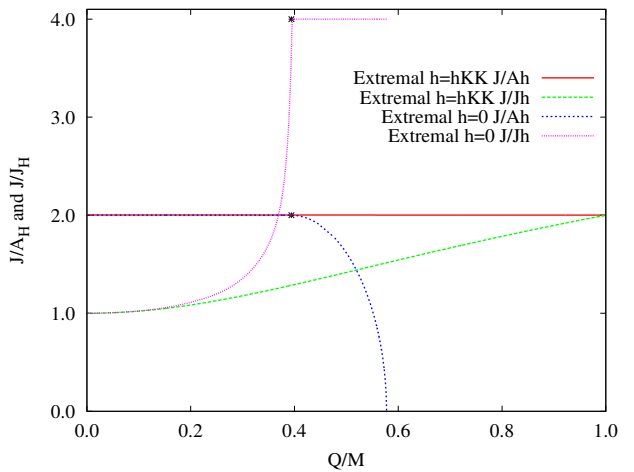}
\end{center}
\caption{\small
The ratios $J/A_{\rm H}$ and
$J/J_{\rm H}$ are shown versus the charge $Q/M$ for extremal $5D$ 
EM ($h=0$) and KK ($h=h_{\rm KK}$) black holes.
The asterisks mark the matching point
of the two EM branches.
}
\label{fig1}
\end{figure}

Fig.~\ref{fig1} exhibits the ratios $J/A_{\rm H}$ and
$J/J_{\rm H}$ versus the charge $Q/M$ for extremal $5D$
EM ($h=0$) and KK ($h=h_{\rm KK}$) black holes.
It clearly reveals the two branches of the
extremal EM solutions, together with their matching point.
This is in constrast to 
the single branch of the EMd solutions, shown
here for the KK case.

\begin{figure}[t!]
\begin{center}
\includegraphics[height=.35\textheight, angle =270]{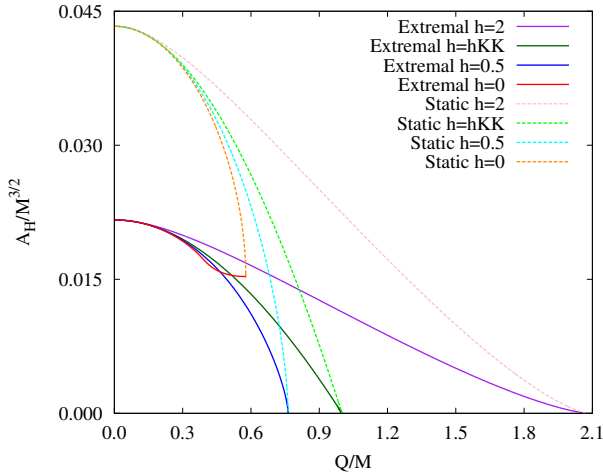}
\end{center}
\caption{\small
The area $A_{\rm H}/M^{3/2}$ is shown versus the charge $Q/M$
for extremal and static $5D$ black holes
for $h=0$, $0.5$, $h_{\rm KK}$ and $2$.
}
\label{fig2}
\end{figure}

We exhibit in Fig.~\ref{fig2} the domain of existence of the
EM and EMd black holes for dilaton coupling constants
$h=0$, 2, 0.5 and $h_{\rm KK}$.
Here we display the area $A_{\rm H}/M^{3/2}$ versus the charge $Q/M$
for extremal and static $5D$ black holes.
All black holes of the respective theories can be found within
these boundaries.
Again we note the different structure for the EM case.
The EM static extremal solution has finite area,
whereas for $h \ne 0$ the static extremal solution is singular
with vanishing area.

\section{EMd black holes in odd $D>5$}

In a straightforward generalization
the near horizon solutions can be constructed for arbitrary odd 
dimensions $D>5$.
In the EM case we retain two branches of solutions, the MP branch
with 
\begin{equation}
J = \sqrt{2(D-3)} A_{\rm H},
\end{equation}
and the Tangherlini branch with
\begin{equation}
J = (D-1) J_{\rm H}.
\end{equation}
In the EMd case the near horizon solutions possess only
a single branch corresponding to the first branch,
with $J = \sqrt{2(D-3)} A_{\rm H}$.

We have performed the respective set of numerical calculations
in $7D$ and in $9D$, and obtained results that are
analogous to the $5D$ case.

\section{Comparison with other theories}

Let us now compare these results with those of two theories whose extremal black holes also exhibit a
branch structure with two distinct branches:
the rotating dyonic black holes of $4$-dimensional KK theory \cite{Rasheed:1995zv},
and the $5$-dimensional black holes of Einstein-Maxwell-Chern-Simons (EMCS)
theory (minimal $D=5$ supergravity) \cite{Chong:2005hr}.

In the first example the $4$-dimensional black holes are characterized by their
mass $M$, angular momentum $J$, electric charge $Q$ and magnetic charge
$P$. In the extremal case, only three of these charges are
independent and two distinct surfaces, \textbf{S}
and \textbf{W}, are found. The restriction to $P=Q$ then yields two distinct branches.
The \textbf{S} branch, $J>PQ$, emerges from the extremal Kerr solution, and
presents all the normal characteristics of charged rotating solutions,
such as an ergo-region and non-zero angular velocity. 
On the other hand, the \textbf{W} branch,
$J<PQ$, possesses no ergo-region and has vanishing horizon angular velocity, although
the angular momentum of the black holes along this branch does not vanish.
At the matching point of both branches,
$J=QP$, the horizon area is zero and the configuration is singular. 

Nevertheless, the
area-angular 
momentum relation for these extremal solutions can be written as
\begin{equation}
A_{\rm H}^2 = 64\pi^2|J^2-Q^2P^2|.
\label{rasheed_area}
\end{equation}
Note, that the electric and magnetic charges are entering the relation for both
branches, and that the 
only difference in the area relation is an overall sign in the expression,
depending on whether we are on the \textbf{S} ($J>PQ$) or on the \textbf{W} ($J<PQ$) branch.

The second example exhibits rather analogous features.
Here we consider
$5$-dimensional black holes in EMCS theory for the supergravity value of the CS
coupling constant, $\lambda=1$ (in an appropriate parametrization). 
In the extremal case, when both angular momenta possess equal magnitude, 
the black holes are parametrized by the angular momentum $J$ 
and the charge $Q$.
Again two branches of extremal black holes are present. 
The first branch has $J^2>-\frac{4}{3\sqrt{3}\pi}Q^3$
and is the ordinary branch with an ergo-region,
while the second branch has $J^2<-\frac{4}{3\sqrt{3}\pi}Q^3$
and is ergo-region free with vanishing horizon angular momentum.
The area-angular momentum
relation for both bran\-ches reads
\begin{equation}
A_{\rm H}^2 = 64\pi^2|J^2 + \frac{4}{3\sqrt{3}\pi}Q^3|.
\label{chong_area}
\end{equation}
At the matching point of both branches the horizon area is again zero and the solution
is singular, and again there is a change of sign in the area-angular momentum
relation depending on the
branch.

Thus in these cases, both charge and angular momentum are entering the area
relation. Moreover, the relations (\ref{rasheed_area}) and (\ref{chong_area}) are in
accordance with the general expressions obtained in \cite{Cvetic:2010mn},
which also depend on both, the charges and the angular momenta.

\section{Further remarks}

It is interesting to note that for the extremal rotating
black holes in EM theory with equal angular momenta, a branch
structure with two distinct branches is found, where for one of the branches - the one emerging from the MP
solution - the area is independent of the charge of the
configuration. 
Along this branch of solutions, the area remains proportional
to the angular momentum and the charge is not entering the relation. 
This is different from other charged black holes considered before.

However, once the critical extremal EM solution\footnote{Note, 
that in contrast to the critical solutions of the $4D$ EMd and
$5D$ EMCS theories discussed above, the critical EM solution is not singular.}
is passed,
the charge enters again into the area relation, yielding the
expression
 \begin{equation}
 A_{\rm H} =  C_1 J^2 Q^{-3/2} + \frac{1}{16 C_1}Q^{3/2},
 \end{equation}
 where $C_1=\frac{3^{1/4}\pi}{\sqrt{2}}$ in our normalization.



In contrast to the two branches of global extremal EM black hole solutions,
the two branches of EM near-horizon solutions do not end at the critical solution.
Thus a study of only  near-horizon solutions is insufficient to clarify
the domain of existence of extremal solutions,
as was first observed for the extremal dyonic black holes of
$D=4$ Gau\ss -Bonnet gravity
\cite{Chen:2008hk}.

Interestingly, in the general EMCS theory (with CS coupling constant $\lambda \ne 1$ 
\cite{Gauntlett:1998fz,Kunz:2005ei}),
there appear even more than two branches of extremal black holes 
for sufficiently large CS coupling
\cite{Blazquez-Salcedo:2013muz}.
As in the case discussed above, however,
the area of these branches of rotating charged black holes 
always depends on both, the charge and the angular momentum. 

Whereas the branch structure of these extremal black holes is very intriguing,
their relation with the corresponding near horizon solutions is surprising as well.
In particular, a given near horizon solution can correspond
to i) more than one global solution, ii) precisely one global solution, or iii) no
global solution at all.
It would be interesting to perform an analogous study
for the general EMd theory (with dilaton coupling constant $h \ne h_{\rm KK}$
\cite{Kleihaus:2003df}),
since the analogy between the known black holes of both theories
suggests that a similar more complex branch structure would be present
for sufficiently large dilaton coupling.

\noindent{\textbf{~~~Acknowledgements.--~}}

We would like to thank B.~Kleihaus and E. Radu for helpful discussions.
We gratefully acknowledge support by the Spanish Ministerio de Ciencia e
Innovacion, research project FIS2011-28013, and by the DFG, in particular, the
DFG Research Training Group 1620 ''Models of Gravity''. J.L.B was supported by
the Spanish Universidad Complutense de Madrid.


\begin{thebibliography}{000}



\bibitem{Tangherlini:1963bw} 
  F.~R.~Tangherlini,
  Nuovo Cim.\  {\bf 27}, 636 (1963).

\bibitem{Myers:1986un}
  R.~C.~Myers and M.~J.~Perry,
  Annals Phys.\  {\bf 172}, 304 (1986).

\bibitem{Aliev:2004ec}
  A.~N.~Aliev and V.~P.~Frolov,
  Phys.\ Rev.\  D {\bf 69}, 084022 (2004)
  [arXiv:hep-th/0401095].

\bibitem{Aliev:2006yk}
  A.~N.~Aliev,
  Phys.\ Rev.\  D {\bf 74}, 024011 (2006)
  [arXiv:hep-th/0604207].

\bibitem{NavarroLerida:2007ez} 
  F.~Navarro-Lerida,
  Gen.\ Rel.\ Grav.\  {\bf 42}, 2891 (2010)
  [arXiv:0706.0591 [hep-th]].

\bibitem{Allahverdizadeh:2010xx} 
  M.~Allahverdizadeh, J.~Kunz and F.~Navarro-Lerida,
  Phys.\ Rev.\ D {\bf 82}, 024030 (2010)
  [arXiv:1004.5050 [gr-qc]].

\bibitem{Allahverdizadeh:2010fn}
  M.~Allahverdizadeh, J.~Kunz and F.~Navarro-Lerida,
  Phys.\ Rev.\ D {\bf 82}, 064034 (2010)
  [arXiv:1007.4250 [gr-qc]].

\bibitem{Kunz:2005nm}
  J.~Kunz, F.~Navarro-Lerida and A.~K.~Petersen,
  Phys.\ Lett.\  B {\bf 614}, 104 (2005)
  [arXiv:gr-qc/0503010].

\bibitem{Kunz:2006eh}
  J.~Kunz, F.~Navarro-Lerida and J.~Viebahn,
  Phys.\ Lett.\  B {\bf 639}, 362 (2006)
  [arXiv:hep-th/0605075].

\bibitem{Kunz:2006jd}
  J.~Kunz, D.~Maison, F.~Navarro-Lerida and J.~Viebahn,
  Phys.\ Lett.\  B {\bf 639}, 95 (2006)
  [arXiv:hep-th/0606005].

\bibitem{Cvetic:2010mn}
  M.~Cvetic, G.~W.~Gibbons and C.~N.~Pope,
  Phys.\ Rev.\ Lett.\  {\bf 106}, 121301 (2011)
  [arXiv:1011.0008 [hep-th]].

\bibitem{Ansorg:2007fh} 
  M.~Ansorg and H.~Pfister,
  Class.\ Quant.\ Grav.\  {\bf 25}, 035009 (2008)
  [arXiv:0708.4196 [gr-qc]].

\bibitem{Hennig:2008yw}
  J.~Hennig, M.~Ansorg and C.~Cederbaum,
  Class.\ Quant.\ Grav.\  {\bf 25}, 162002 (2008)
  [arXiv:0805.4320 [gr-qc]].

\bibitem{Ansorg:2008bv} 
  M.~Ansorg and J.~Hennig,
  Class.\ Quant.\ Grav.\  {\bf 25}, 222001 (2008)
  [arXiv:0810.3998 [gr-qc]].

\bibitem{Hennig:2008zy} 
  J.~Hennig, C.~Cederbaum and M.~Ansorg,
  Commun.\ Math.\ Phys.\  {\bf 293}, 449 (2010)
  [arXiv:0812.2811 [gr-qc]].

\bibitem{Ansorg:2009yi} 
  M.~Ansorg and J.~Hennig,
  Phys.\ Rev.\ Lett.\  {\bf 102}, 221102 (2009)
  [arXiv:0903.5405 [gr-qc]].

\bibitem{Hennig:2009aa} 
  J.~Hennig and M.~Ansorg,
  Annales Henri Poincare {\bf 10}, 1075 (2009)
  [arXiv:0904.2071 [gr-qc]].

\bibitem{Ansorg:2010ru}
  M.~Ansorg, J.~Hennig and C.~Cederbaum,
  Gen.\ Rel.\ Grav.\  {\bf 43}, 1205 (2011)
  [arXiv:1005.3128 [gr-qc]].


\bibitem{Castro:2012av} 
  A.~Castro and M.~J.~Rodriguez,
  Phys.\ Rev.\ D {\bf 86}, 024008 (2012)
  [arXiv:1204.1284 [hep-th]].

\bibitem{Astefanesei:2006dd}
  D.~Astefanesei, K.~Goldstein, R.~P.~Jena, A.~Sen and S.~P.~Trivedi,
  JHEP {\bf 0610} (2006) 058
  [arXiv:hep-th/0606244].

\bibitem{Goldstein:2007km}
  K.~Goldstein and R.~P.~Jena,
  JHEP {\bf 0711} (2007) 049
  [arXiv:hep-th/0701221].

\bibitem{Kunduri:2013gce}
  H.~K.~Kunduri and J.~Lucietti,
  arXiv:1306.2517 [hep-th].


\bibitem{COLSYS}
 U. Ascher, J. Christiansen, R.~D. Russell,
 Mathematics of Computation 33 (1979) 659;
 ACM Transactions 7 (1981) 209.




\bibitem{Rasheed:1995zv}
  D.~Rasheed,
  Nucl.\ Phys.\  B {\bf 454}, 379 (1995)
  [arXiv:hep-th/9505038].


\bibitem{Chong:2005hr}
  Z.~W.~Chong, M.~Cvetic, H.~Lu and C.~N.~Pope,
  Phys.\ Rev.\ Lett.\  {\bf 95}, 161301 (2005)
  [arXiv:hep-th/0506029].

\bibitem{Chen:2008hk} 
  C.~-M.~Chen, D.~V.~Gal'tsov and D.~G.~Orlov,
  Phys.\ Rev.\ D {\bf 78}, 104013 (2008)
  [arXiv:0809.1720 [hep-th]].


\bibitem{Gauntlett:1998fz} 
  J.~P.~Gauntlett, R.~C.~Myers and P.~K.~Townsend,
  Class.\ Quant.\ Grav.\  {\bf 16}, 1 (1999)
  [hep-th/9810204].

\bibitem{Kunz:2005ei} 
  J.~Kunz and F.~Navarro-Lerida,
  Phys.\ Rev.\ Lett.\  {\bf 96}, 081101 (2006)
  [hep-th/0510250].


\bibitem{Blazquez-Salcedo:2013muz}
  J.~L.~Blazquez-Salcedo, J.~Kunz, F.~Navarro-Lerida and E.~Radu,
  arXiv:1308.0548 [gr-qc].

\bibitem{Kleihaus:2003df} 
  B.~Kleihaus, J.~Kunz and F.~Navarro-Lerida,
  Phys.\ Rev.\ D {\bf 69}, 081501 (2004)
  [gr-qc/0309082].

\end{thebibliography}
\end{document}